\documentclass[runningheads]{llncs}
\usepackage{graphicx} 
\usepackage{amsfonts}
\usepackage[T1]{fontenc}
\usepackage{amsmath,amssymb,amsfonts}
\usepackage{cite}
\usepackage{tikz}
\usetikzlibrary{quantikz2}
\usepackage{physics}
\graphicspath{ {./images/} }
\usepackage{algorithmic}
\usepackage{fancyhdr}
\usepackage{microtype}
\usepackage{subfigure}
\usepackage{booktabs}
\usepackage{hyperref}

\usepackage[colorinlistoftodos,prependcaption,textsize=small]{todonotes}
\definecolor{rev1color}{RGB}{220,100,0}    
\definecolor{rev2color}{RGB}{0,80,180}     
\definecolor{rev3color}{RGB}{130,0,130}    

\usepackage{acronym}
\newacro{AUPRC}{Area Under Precision-Recall Curve}
\newacro{QML}{Quantum Machine Learning}
\newacro{PQC}{Parameterized Quantum Circuits}
\newacro{VQC}{Variational Quantum Circuit}
\newacro{QNN}{Quantum Neural Network}
\newacro{GQC}{Guided Quantum Compression}
\newacro{UMAP}{Uniform Manifold Approximation and Projection}
\newacro{PCA}{Principal Component Analysis}
\newacro{FNN}{Feedforward Neural Network}
\newacro{MSE}{Mean Squared Error}
\newacro{IQP}{Instantaneous Quantum Polynomial}

\date{\today}

\begin{document}

\title{A Mixture-of-Experts Framework for Practical Hybrid-Quantum Models in Credit Card Fraud Detection}
\titlerunning{MoE Framework for Hybrid-Quantum Models in Fraud Detection}






\author{Rodrigo Chaves\inst{1} \and
Kunal Kumar\inst{2} \and
Bruno Chagas\inst{3} \and
Rory Linerud\inst{4} \and
Brannen Sorem\inst{4} \and
Javier Mancilla\inst{1} \and
Bryn Bell\inst{1}}
\authorrunning{R. Chaves et al.}

\institute{Hardware Engineering, Oxford Quantum Circuits, United Kingdom\\
\email{\{rchaves, jmancilla, bbell\}@oqc.tech} \and
Hardware Engineering, Oxford Quantum Circuits, Japan\\
\email{kkumar@oqc.tech} \and
Foundry, Mastercard, Ireland\\
\email{bruno.chagas@mastercard.com} \and
Foundry, Mastercard, United States of America\\
\email{\{rory.linerud, brannen.sorem\}@mastercard.com}
}

\maketitle

\begin{abstract}
    This paper investigates whether hybrid quantum–classical machine learning can deliver practical improvements in financial fraud detection performance for card-based and other payment transactions. Building on a Guided Quantum Compressor architecture, the approach integrates an autoencoder, a variational quantum circuit, and a classical neural head, and then embeds this hybrid model into a mixture‑of‑experts framework including a state‑of‑the‑art gradient‑boosted tree classifier. Using a European credit card dataset with severe class imbalance, the routed hybrid architecture with 0.6 threshold achieves average precision scores of $0.793\pm0.085$ compared to $0.770\pm0.096$ of XGBoost on 3 repeated 5-fold cross-validation benchmarks. Precision and recall comparisons reveals a possible trade-off of fraud and nominal detections with a reduction in false positives at the cost of a small reduction in fraud detections. The improvements are achieved while adding only 7 to 21 minutes of extra inference time depending on the choice of hyperparameters. These results indicate that selectively routing transactions to quantum–classical models can enhance fraud detection while remaining compatible with the latency and operational constraints of modern financial institutions.
\end{abstract}



\section{Introduction}
Financial fraud remains a critical global challenge, eroding trust in institutions, destabilizing markets, and generating substantial economic losses \cite{Cao2021}. Across sectors—including banking, insurance, and investment—organisations face increasingly sophisticated fraud schemes, insider threats, and external attacks \cite{Nicholls2021}. The rapid growth and complexity of financial data, spanning online transactions, credit operations, and cross-border flows, challenge traditional manual or rule-based detection methods \cite{Zhu2024}. Online payment fraud alone is projected to cost merchants over \$343 billion globally between 2023 and 2027 \cite{JuniperResearch2023}, while surveys indicate that the majority of organisations continue to experience attempted or actual fraud, highlighting the persistent and evolving nature of the threat \cite{JPmorgan2022}.

Fraud occurs in various forms, such as credit card fraud, loan default schemes, money laundering, insurance fraud, and fraudulent financial statements. Despite this diversity, external fraud dominates the detection literature - especially in credit, loan, and insurance contexts \cite{Wickramanayake2020}. Identifying such fraud remains a tough challenge because fraudulent cases are extremely rare, resulting in severe class imbalance \cite{Baisholan2025}. Moreover, adversarial behavior continuously evolves, requiring detection systems to adapt to changing strategies and emerging attack patterns \cite{AlDaoud2025}.

In recent years, machine learning has emerged as the primary approach for detecting financial fraud, and often leverages supervised classification, unsupervised anomaly detection, and semi-supervised or reinforcement learning techniques \cite{Compagnino2025}. These methods support automated identification of patterns, anomalies, and abnormal behaviours within large, high-dimensional transaction and financial datasets, surpassing the capabilities of manual analysis \cite{Ali2022}. Though traditional models such as decision trees, support vector machines, and random forests remain prevalent, advanced approaches including deep learning architectures such as autoencoders and long short-term memory networks, as well as graph-based models, are increasingly used in contemporary fraud detection systems \cite{Compagnino2025}.

Fraudulent transactions are rare and continuously evolving, making accurate detection highly challenging. Financial transaction data is high-dimensional and complex, encompassing temporal, relational, and heterogeneous features, which can limit the effectiveness of classical machine learning models \cite{zaffar2023,Breskuvien2024}. Traditional approaches are often vulnerable to adversarial attacks, where fraudsters manipulate their behavior to evade detection \cite{AlDaoud2025}. At the same time, interpretability and regulatory compliance remain essential, as financial institutions must explain detection decisions to both regulators and customers to maintain trust \cite{Aljunaid2025, erneviien2024}. The consequences of detection errors are substantial: false declines are projected to cost merchants \$430 billion globally - approximately 75 times higher than actual fraud losses \cite{FICO2024} - and around 1 in 15 consumers are affected by fraud-driven declines \cite{Javelin2018}. Nearly 40\% of cardholders abandon their card after a false decline, and 20\% of cards show no activity for six months following such events \cite{Javelin2015}. These findings underscore the urgent need for more accurate and robust fraud detection systems.

Quantum models such as variational quantum circuits and quantum kernel methods demonstrate potential to capture complex structures in transaction data and improve the detection of rare fraud cases \cite{Grossi2022,Kyriienko2022}. Reference \cite{Grossi2022} in particular identifies complementarity between quantum and classical methods as a potential source of advantage in binary classification and fraud detection problems. Specifically, it is emphasized that quantum and classical models could detect different relationships or structures in the data, and that this difference could itself be exploited in a hybrid setting. By enriching feature representations, quantum algorithms may achieve stronger separation between legitimate and fraudulent behaviors \cite{Havlek2019}. Quantum machine learning models also offer advantages in generalisation and the efficient handling of high-dimensional feature spaces, which can lead to more expressive and robust decision boundaries \cite{Caro2022, Huang2021, Biamonte2017}. Furthermore, hybrid quantum and classical frameworks present a promising direction for scalable fraud detection systems that can adapt to evolving attack strategies while potentially enabling faster processing over large datasets \cite{Cardaioli2025, Ubale2025}.

Our main contributions are two. First, an improved version of the Guided Quantum Compressor model where we introduce a classification head to provide a confidence interval, use a different ansatz from the initially proposed, and calibrate the confidence interval to achieve better results in credit card fraud detections with details found in section \ref{sec:overview}. The second contribution is the use of a quantum-classical mixture-of-experts to achieve a method for low-latency classification with quantum-classical hybrids in quantum devices with low clock speeds that achieves results comparable to models like XGBoost.



Section~\ref{sec:experiments} reports our empirical results using a publicly available European credit card fraud dataset, including the complete pipeline from data preprocessing to model training and evaluation. Finally, Section~\ref{sec:discussion} discusses our findings, outlines current limitations, and highlights open research directions for future work.


\section{Theoretical background and methodology}\label{sec:overview}

Credit card fraud detection is a critical application of machine learning, characterized by extreme class imbalance, high-dimensional time series data, and adversarial non-stationarity. Fraudulent transactions typically account for far less than 1\% of all transactions \cite{LeBorgne2022}, making it challenging to train models that generalize well. Moreover, fraudsters continuously evolve their methods, so models must adapt quickly through frequent retraining or online learning. Latency constraints further complicate deployment: transactions must be classified in near real-time to prevent financial loss, making performance at inference as important as predictive accuracy.

Effective fraud detection systems require a careful balance of precision and recall, robust handling of data imbalance, real-time inference capabilities, and resilience against security threats \cite{DalPozzolo2014, LeBorgne2022}. Accuracy alone is not a suitable metric in this context due to the heavy skew of the dataset toward legitimate transactions; metrics such as F1-score, precision-recall curves, and area under the precision-recall curve (AUPRC) are preferred. 


Quantum machine learning (QML) offers complementary tools for tackling these challenges. Variational quantum machine learning uses parametrized quantum circuits (PQCs), where tunable angles are optimized via classical algorithms \cite{benedetti2019pqc}. Stacking these circuits into layers gives rise to quantum neural networks (QNNs) \cite{beer2020qnn}, which are well-suited for modeling intricate relationships in financial data \cite{doosti2024}. Quantum kernel methods provide an alternative by embedding data into a high-dimensional quantum feature space and computing similarities between points for downstream classical classifiers, allowing subtle structures in financial transactions to be more effectively captured \cite{schuld2019quantumkernel}.

Recent studies show that QML could enhance fraud detection. Resource-conscious variational quantum models optimize feature embeddings while minimizing qubit requirements and circuit depth \cite{Das2025}. Comparative analyses suggest some quantum classifiers show promise of outperforming classical methods under certain scenarios \cite{Grossi2022}, and hybrid architectures that integrate quantum circuits with recurrent neural networks could improve temporal modelling of transaction sequences \cite{Ubale2025}. Quantum kernel methods have also been successfully applied to anomaly detection under hardware noisy environments \cite{pranjic2024}, while quantum graph-based models exploit relational structures inherent in financial networks \cite{Innan2024}. Together, these advances indicate that quantum techniques, either stand-alone or in hybrid configurations, may complement classical methods to achieve higher accuracy, better generalization, and efficient real-time detection, though practical deployment remains constrained by current hardware and scalability limitations.


\subsection{Quantum Model}
Our model builds upon the Guided Quantum Compressor (GQC) architecture originally proposed by Belis \textit{et al.} \cite{Belis2024}. We extend their framework to create a custom hybrid quantum-classical architecture, which consists of the following key components: dimensionality reduction of the input data using a trained autoencoder; mapping of the reduced features into a quantum space via angle encoding; classification using a variational quantum circuit (VQC); a classical feedforward neural network serving as the final classification head; and calibration of the predicted outputs using temperature scaling based on Platt’s method.

Each of these components is described in detail in the following subsections, along with the main modifications we introduced relative to the original GQC architecture.

Similarly to the base architecture, we also perform epoch training using a linear combination of the loss associated with the reconstruction of the autoencoder, $\mathcal{L}_R$, and the loss of the classification, $\mathcal{L}_C$. Our aim is then to minimize the following total loss, 
\begin{equation}
    \mathcal{L}_{total} = \lambda\mathcal{L}_R + (1-\lambda)\mathcal{L}_C,
\end{equation}
where $\lambda\in[0,1]$. Further details of each component are provided in the following sections.

\subsubsection{Motivation} Fraud detection is considered a very low-latency application of machine learning with the requirements that the client must have the approval (or rejection) decision ideally within milliseconds and at most, in special cases, a 3 second window. This was our main motivation to pursue the Mixture-of-Experts strategy outlined later in this paper. To select the most appropriate quantum model for the task, the initial benchmark performed was on the required inference time of three different quantum models: Quantum Multiple Kernel Learning (QMKL) \cite{Miyabe2023}, the chosen Guided Quantum Compressor (GQC) \cite{Belis2024}, and a Genetic Feature Map \cite{Tjandra2024}.

Average inference time per sample served as the primary benchmark for model selection. To evaluate this, we reduced the proposed dataset to 10,000 samples, applied a 90/10 train-test split, balanced the training set to 90 samples per class, and recorded the post-training inference time. Given that the experiments were restricted to two qubits, the kernel methods required PCA for dimensionality reduction; in contrast, GQC integrates this reduction directly into its architecture. All evaluations were performed exclusively in a CPU environment running PennyLane 0.41.1, PyTorch 2.7.0, and NumPy 1.26.4. The benchmarking protocol was structured into five timing runs, with each run executing 10 internal repetitions on a batch of 50 samples. The initial run served solely as a hardware warm-up and was excluded from the final measurements.

Table \ref{tab:results-ait} presents the average inference time per sample (in milliseconds) and the coefficient of variation (CV) for the evaluated algorithms. Inference latency spans more than three orders of magnitude across the three methods. Specifically, GQC operates approximately 542× faster than GFM and 1,387× faster than QMKL per sample. While GFM and QMKL are both kernel methods, GFM is approximately 2.6× faster because it evaluates a single evolved kernel per (test, train) pair, whereas QMKL evaluates three. Absolute standard deviation is the critical metric for Service Level Agreements (SLAs), and GQC improves upon the kernel methods by two orders of magnitude (0.004 ms compared to 0.189–0.423 ms). Furthermore, GQC's higher CV (4.25\%) is a measurement artifact rather than an indicator of instability; with a per-sample mean of approximately 89 µs, the execution time approaches the system's timer-resolution and OS-scheduling noise floor, causing minor absolute jitter to appear proportionally large. Conversely, the kernel methods exhibit CVs under 0.4\% simply due to their substantially larger absolute means, not due to greater determinism.

The inner repetition loop within the benchmarking protocol was necessary to render GQC measurable at all. Under the latency criterion alone, only GQC is Pareto-optimal. Both GFM and QMKL are strictly dominated, as GQC is not only significantly faster but also incidentally achieves a higher AUC. Even if the AUC dimension is excluded, neither kernel method provides a latency advantage in any regime; there is no operating point where selecting GFM or QMKL over GQC yields a beneficial trade-off involving latency. One point to notice is that GFM and QMKL inference time grows $\mathcal{O}(N)$ and $\mathcal{O}(K*N)$, respectively, where $N$ is the number of training samples and $K$ the number of kernels. Although the inference time looks achievable in the context of low-latency scenarios, it needs to be observed that only 180 training samples were used in this benchmark.

\begin{table}[ht]
    \centering
    \caption{Table containing the average inference time per sample in milliseconds and coefficient of variation of QMKL, GQC, and Genetic Feature Map.}
\resizebox{0.7\linewidth}{!}{%
\begin{tabular}{ |p{3cm}||p{3cm}|p{3cm}|p{3cm}|  }
 \hline
 Algorithm & Inference (ms/sample)  & CV \% \\
 \hline
 QMKL   & $123.9\pm0.4$ & 0.34\\
 Genetic Feature Map&   $48.4\pm0.2$  & 0.39\\
 GQC &$\mathbf{0.089\pm0.004}$ & 4.25\\

 \hline
\end{tabular}}

    \label{tab:results-ait}
\end{table}

\subsubsection{Autoencoders}
Due to the limitation of current hardware, a dimensionality reduction or feature selection strategy is often required. Different strategies are available in the machine learning literature ranging from UMAP \cite{McInnes2018}, tree-based feature selections \cite{Breiman2001}, PCA \cite{Hotelling1936, Pearson1901}, and others, each case having their pros and cons \cite{Huang2022, Armostrong2022, Barbieri2024}. Noticing that quantum computing relies on the entanglement and correlation of features to obtain a better classification \cite{Sharma2022}, it is not yet clear what type of dimensionality reduction or feature selection is best suited for quantum computers. Autoencoders were first proposed in the context of unsupervised learning due to their ability to learn important features by trying to recreate the original data \cite{Hinton2006}.

A feedforward neural net (FNN) with ReLU activation functions is used to encode the data in a lower latent dimension of the data and another FNN is used to decode the data back to its original form. A loss function, $\mathcal{L}_R$, is then defined to train the model to ensure that the original data can be reconstructed from the encoded one, and MSE (mean-squared error) loss function is used for this:
\begin{equation}
    \mathcal{L}_R = \frac{1}{n}\sum_{i=1}^{n}(\textbf{x}_i-\textbf{y}_i)^2
\end{equation}
where $n$ is the number of inputs, $\textbf{y}_i$ is the i-th reconstructed vector, and $\textbf{x}_i$ is the i-th vector of the input.

The main difference from the base GQC model comes from the input data of the autoencoder. In our work, we only train the autoencoder with non-fraudulent cases instead of all possible cases. This is to guarantee that the reduced input has as much information as possible about the non-fraudulent cases when passed to our variational classifier.

\subsubsection{Angle encoding}
Although quantum computers provide a natural way of working with higher-dimensional Hilbert spaces, they still inherently operate within quantum systems. Data must be represented in a form readable by quantum computers, and quantum mechanics imposes known limitations associated with the linear spaces they can represent. Since classical data are not subject to the same constraints, one of the open challenges in quantum machine learning is finding an effective way to encode classical data within a quantum mechanical framework.

There are multiple strategies for encoding classical data into quantum computers, including amplitude encoding, phase encoding, Z feature maps, ZZ feature maps, and others.

In this work, we chose angle encoding as our encoding method. Angle encoding is widely used in various QML models such as Dressed Quantum Circuits \cite{Mari2020}, Quantum Kitchen Sinks \cite{Wilson2019}, IQP Variational Classifiers \cite{Havlíček2019}, and many others. A general one-qubit state $\ket{\psi}$ can be expressed as:
\begin{equation}
    \ket{\psi} = \cos(\frac{\theta}{2})\ket{0} + e^{i\phi}\sin(\frac{\theta}{2})\ket{1}.
\end{equation}
Angle encoding uses the angle $\theta$ defined by rotations on an axis of the qubit, such as X or Y, to encode the features of the data.

As an example, assume that we are encoding the k-th feature of the j-th data point in our data set, $\textbf{x}_k^j$ using Y-axis rotations. Then the final state representing this feature value is:
\begin{equation}
    \ket{x_k^j} = RY(\theta=x_k^j)\ket{0} = \cos\bigg(\frac{\textbf{x}_k^j}{2}\bigg)\ket{0}+\sin\bigg(\frac{\textbf{x}_k^j}{2}\bigg)\ket{1}.
\end{equation}
As shown, this encoding method requires $n$ qubits for $N$ features. However, alternative formulations such as dense angle encoding \cite{LaRose2020} allow multiple features to be encoded within a single qubit, though this approach was not used in our work.

It is also worth noting that the general qubit state includes a complex phase term $e^{i\phi}$. In the context of angle encoding, such complex phases naturally appear when rotations around the X-axis are performed instead of the Y-axis. However, there are multiple ways to encode data in quantum computers and different methods lead to different decision boundaries in the quantum machine learning model \cite{LaRose2020, Schuld2018}. 

In this work, we used Y-axis rotations for angle embedding.

\subsubsection{Variational Quantum Circuit (VQC) Classifier}
The variational quantum circuits (VQC) is a quantum machine learning model where a circuit is constructed with parametrized rotations that are trained variationally to perform classification \cite{Mari2020, Wilson2019, Havlíček2019, Huang2021}, regression \cite{Salinas2021}, or generative modelling \cite{Barthe2025}. This is a building block of what was labelled as quantum neural network models.

The VQC can be expressed as:
\begin{equation}
    \mathcal{C}(\mathbf{\theta}, \mathbf{x}) = \Tr[\mathcal{M}\rho(\mathbf{\theta}, \mathbf{x})],
\end{equation}
where $\mathcal{M}$ is an operator and $\rho$ a density matrix where both depends on trainable parameters $\mathbf{\theta}$ and input data $\mathbf{x}$. Usually, a VQC model will be composed of one or more variational circuits and it will be equipped with a prediction function such that
\begin{equation}
    y = f_{\text{pred}}(\mathcal{C}(\mathbf{\theta}, \mathbf{x}))
\end{equation}
where $y\in\{+1, -1\}$ represents the predicted label.


In our case, the prediction function is replaced by a classical feedforward neural network, which will be described in the following section.

When designing the quantum circuit ansatz, two considerations must be taken into account: trainability and expressibility. The former is related to the gradient-vanishing problem in variational quantum circuits. During training, gradients may vanish, preventing meaningful improvements in the loss function across epochs. This phenomenon, known as the barren plateau problem, is influenced by the circuit depth, the choice of gates, and the cost function $\mathcal{C}$ chosen for the VQC \cite{Cerezo2021, Larocca2025}. Expressibility, on the other hand, refers to the diversity of states produced by the circuit. Higher expressibility means that the output probability distributions more closely approximate the $t$-moment of the Haar measure on $SU(n)$ \cite{Cerezo2021}.

Although it might seem intuitive that higher expressibility would lead to better performance in quantum machine learning models, there exists a trade-off between trainability and expressibility \cite{Larocca2025}. Balancing these two factors is crucial to ensure that the model achieves both effective optimization and sufficient representational power without falling into potential pitfalls.

The ansatz chosen by us is called Alternating Layered Ansatz and it consists of $n$ layers of Y-axis and Z-axis rotations followed by nearest-neighbours CNOT gates. To guarantee trainability of this ansatz, we use the expectation value of a single qubit Pauli-Z operator as the cost function $\mathcal{C}$ \cite{Cerezo2021}.

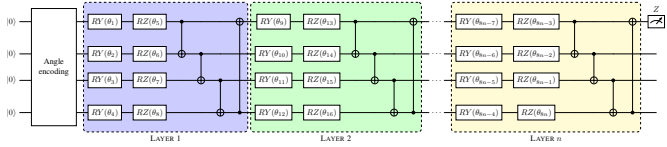
\begin{figure}
\resizebox{\linewidth}{!}{%
\begin{quantikz} 
\lstick{$\ket{0}$} & \gate[4]{\begin{array}{c}\text{Angle}\\\text{encoding}\end{array}} &  \gate{RY(\theta_1)}\gategroup[4,steps=6,style={dashed,rounded corners,fill=blue!20, inner xsep=2pt},background,label style={label position=below,anchor=north,yshift=-0.2cm}]{{\sc Layer 1}} & \gate{RZ(\theta_5)} & \ctrl{1} & & & \targ{}&\gate{RY(\theta_9)}\gategroup[4,steps=6,style={dashed,rounded corners,fill=green!20, inner xsep=2pt},background,label style={label position=below,anchor=north,yshift=-0.2cm}]{{\sc Layer 2}} &  \gate{RZ(\theta_{13})} & \ctrl{1} & & & \targ{} & \ \ldots\ & \gate{RY(\theta_{8n-7})}\gategroup[4,steps=6,style={dashed,rounded corners,fill=yellow!20, inner xsep=2pt},background,label style={label position=below,anchor=north,yshift=-0.2cm}]{{\sc Layer $n$}} & \gate{RZ(\theta_{8n-3})} & \ctrl{1} & & & \targ{} & \meter{Z}\\ 
\lstick{$\ket{0}$} & & \gate{RY(\theta_2)} & \gate{RZ(\theta_6)} & \targ{} & \ctrl{1} & &&\gate{RY(\theta_{10})} & \gate{RZ(\theta_{14})} & \targ{} & \ctrl{1} & & & \ \ldots\ & \gate{RY(\theta_{8n-6})} & \gate{RZ(\theta_{8n-2})} & \targ{} & \ctrl{1} & &&\\ 
\lstick{$\ket{0}$} & & \gate{RY(\theta_3)} & \gate{RZ(\theta_7)} & & \targ{} & \ctrl{1} &&\gate{RY(\theta_{11})} & \gate{RZ(\theta_{15})} & & \targ{} & \ctrl{1} & & \ \ldots\ & \gate{RY(\theta_{8n-5})} & \gate{RZ(\theta_{8n-1})} & & \targ{} & \ctrl{1} &&\\ 
\lstick{$\ket{0}$} & & \gate{RY(\theta_4)} & \gate{RZ(\theta_8)} & & & \targ{} & \ctrl{-3} &\gate{RY(\theta_{12})} & \gate{RZ(\theta_{16})} & & & \targ{} & \ctrl{-3} & \ \ldots\ & \gate{RY(\theta_{8n-4})} & \gate{RZ(\theta_{8n})} & & & \targ{} & \ctrl{-3} &
\end{quantikz}
}%
    \caption{Circuit used in the architecture as the classifier for 4 qubits with $n$ layers.}
    \label{fig:quantum_classifier}
\end{figure}

The main difference from the base architecture for GQC is that their variational circuits interleaves data uploading layers with variational layers with each possibly happening more than once. In our work, the data encoding step occurs only once. An example of what the circuit would be for four qubits can be found in figure \ref{fig:quantum_classifier}.

\subsubsection{Feedforward Neural Networks}

One of the main differences between our proposed GQC architecture and the original design lies in the way that the model produces label predictions. In the original GQC, this is done by applying a sign function such that the label is given by the prediction function \cite{Belis2024}
\begin{equation}
   f_{pred}(\mathcal{C}(\mathbf{\theta}, \mathbf{x})) =  \frac{\text{sign}[\mathcal{C}(\mathbf{\theta}, \mathbf{x})]+1}{2},
\end{equation}
where $\mathcal{C}(\mathbf{\theta}, \mathbf{x})$ is the cost function defined by an operator.

In our work, we introduce a classification head with two goals. First, it enables the possibility of fine-tuning the backbone composed of the quantum model. Second, to introduce a confidence interval for the classification. The concept of using a classification head is not unfamiliar in the AI literature and different heads were used in image classification \cite{Alex2017}, large language models \cite{Devlin2019}, and multi-label classification \cite{Coope2019}, and many other domains. Each architecture employs different heads (or multiple heads) to achieve better performance and, in our case, we use a simple feedforward neural network (FNN).

FNNs were among the first proposed architectures in deep learning. They follow a straightforward logic in which information flows from input to output without cycles in the intermediate layers. They have been applied in a wide range of settings, including pattern recognition \cite{Jain2000}, clustering and classification \cite{Zhang2000}, and bioinformatics \cite{Mitra2006}. FNNs are well known for their ability to identify non-linearly separable patterns \cite{Rumelhart1986} and to approximate any continuous function \cite{Hornik1991}. In our case, we use a special case of FNNs called multilayer perceptron (MLP) which consists of fully connected nodes in each layer and non-linear activation functions. Similar to the autoencoders, we use ReLU activation functions in our classification head.

As the classification head, a loss function for classification, $\mathcal{L}_C$, is defined to perform epoch training. In our case, we use a BCELoss defined by the mean of the Binary Cross Entropy between the ground truth and the predicted probabilities. This is defined as:
\begin{equation}
    \mathcal{L}_C = -\frac{1}{n}\sum_{i=1}^{n}y_i\log(\hat{p}_i) + (1-y_i)\log(1-\hat{p}_i), 
\end{equation}
where $n$ is the total number of inputs, $\textbf{y}$ is the ground truth labels, and $\hat{\textbf{p}}$ the predicted probabilities of the classification head.

As mentioned earlier, both the classification head and the autoencoder in our architecture use MLPs with varying depths and numbers of nodes. Together, these components provide a full description of our complete quantum–classical hybrid architecture.

\subsubsection{Temperature calibration}
\label{subsubsec:temp_calib}
One final step of the architecture is calibration. Modern neural networks are often miscalibrated, meaning that their predicted confidence values do not accurately reflect the true probability of correctness \cite{Guo2017}. In mathematical terms, define random variables for the input $x \in \mathcal{X}$
and the label $y\in\mathcal{Y}=\{0,1\}$ that follows the following joint distribution:
\begin{equation}
    \mathbf{P}(x,y) = \mathbf{P}(y|x)\mathbf{P}(x).
\end{equation}
Let $h(x)=(\hat{y}, \hat{p})$ be a neural network where $\hat{y}$ is the class prediction and $\hat{p}$ is the probability of correctness. Perfect calibration is defined as
\begin{equation}
    \mathbf{P}(\hat{y} = y | \hat{p} = p) = p,\qquad \forall p\in[0,1].
\end{equation}
What this means is that we want the probability of correctness to represent a real probability. For example, if we sample 100 predictions for which the model assigns a confidence of 0.6, then approximately 60 of them should be correct. Since the sampling process is discrete while the random variable $\hat{p}$ is continuous, this means that perfect calibration can only be approximated in practice.

In this work, we use a variant of Platt scaling \cite{Platt2000}, called temperature calibration, consistent with the findings of Guo \textit{et al.} \cite{Guo2017}, who showed that it is an effective calibration method in general settings. Temperature calibration consists of obtaining the log-odds of the predicted probability of an evaluation set defined as
\begin{equation}
    \text{logits}(\hat{p}_i) = \ln\bigg(\frac{\hat{p}_i}{1-\hat{p}_i}\bigg).
\end{equation}
Then a scaled probability, $\hat{\pi}$, is obtained from the logits
\begin{equation}
    \hat{\pi}_{i} = \frac{1}{1+\exp(\frac{-\text{logits}(\hat{p}_i)}{t})},
\end{equation}
where $t$ is a temperature parameter that is going to be minimized according to a negative likelihood loss function (NLL loss), $\mathcal{L}_{nll}$, defined by
\begin{equation}
    \mathcal{L}_{nll} = -\frac{1}{n}\sum_{i=1}^{n}y_i\log(\hat{\pi}_{i}) + (1-y_i)\log(1-\hat{\pi}_{i}),
\end{equation}
where $n$ is the total number of inputs, $y$ the ground truth label, and $\hat{\pi}$ the scaled predicted correctness.


\subsection{Combined Model}
\label{subsec:combined_model}

Ensemble learning \cite{Zhou2025} is a widely used technique in machine learning  to improve the classification of weak learners. Methods like bagging \cite{Efron1992}, boosting \cite{Schapire1990, Mason1999}, and stacking \cite{Smyth1997} are very well-known with the first two being fundamental pieces of models like XGBoost \cite{chen2016xgboost} and CatBoost \cite{prokhorenkova2018catboost} which are the current state-of-the-art (SOTA) in classification. 

In this work, we focus on the mixture-of-experts (MoE), another ensemble learning technique, for two reasons.
First, during our experiments, we observed that the correctly classified instances produced by our architecture did not fully intersect with those correctly classified by SOTA models such as XGBoost.
Second, one of the major challenges in adopting current QML models is the latency associated with quantum hardware. In the context of fraud detection for financial institutions, even strategies such as train on classical, deploy on quantum \cite{Recio2025} are infeasible due to the extremely low-latency requirements of financial transactions (e.g., credit card authorization).

Our aim is to develop a method that pre-selects the most favourable cases for quantum inference, improving the overall performance of SOTA classical methods while keeping latency competitive for real-world scenarios.

\subsubsection{Mixture-of-experts }

Mixture-of-Experts (MoE) \cite{Masoudnia2014, Cai2025} was first introduced by Jacobs et al. in two seminal papers \cite{Jacobs1991, jordan1994hme}. It was originally proposed as a “divide-and-conquer” learning strategy that differs from traditional dense models. Initially developed for neural networks, MoE was designed to avoid activating all the parameters of a neural network for every input. Instead, the model enables the dynamic selection of parameters based on their relevance. Theoretical and experimental work has demonstrated a connection between the effectiveness of the combining procedure and the correlation of errors among the experts \cite{Tumer1996}, with the best performance occurring when the errors are negatively correlated. Its use has been wildly spread with applications in image classification \cite{Riquelme2021}, image generation \cite{Xue2023, Park2018}, semantic segmentation \cite{Wang2020}, and, more notably, in large language models with the DeepSeek model \cite{deepseekai2024}.

Central to the design of MoE is its routing mechanism. As first proposed by Jacobs \cite{Jacobs1991}, the input space is partitioned into subspaces by a gating or routing system that is trained separately from the experts. In this approach, rather than computing the predictions of all experts and combining them using some weighted function as in stacking \cite{Smyth1997}, the trained router dynamically selects the most appropriate expert based on the given input.

Building on the classical formulation of MoE, our work adapts the routing and expert method to a hybrid quantum-classical architecture. Other than relying on dense evaluation of the quantum model, we propose a method to enhance SOTA algorithms, in this case XGBoost, by using a third model as the routing mechanism in the architecture. As mentioned above, this method is important to maintain competitive latencies to current classical models that are required in real-world settings such as fraud detection. By integrating a MoE-like router with quantum variational classifier and a classical model, our approach aims to exploit the complementary strengths of both paradigms while maintaining practical computational efficiency.

\subsubsection{Experts training}

The combined architecture takes as input two base classifiers
\(
h^{(1)}, h^{(2)} : \mathcal{X} \to [0,1]
\),
and a router (or gating) model
\(g : \mathcal{X} \to [0,1]\).
In our experiments, the \emph{primary} model \(h^{(1)}\) is a strong gradient-boosted tree learner such as XGBoost~\cite{chen2016xgboost} or CatBoost~\cite{prokhorenkova2018catboost}, while the \emph{secondary} model \(h^{(2)}\) is the quantum–classical hybrid architecture described above. Both base models are wrapped in a temperature-calibration layer (Section~\ref{subsubsec:temp_calib}), so that they output calibrated class probabilities

\[
\hat{\pi}^{(1)}(x) = \mathbf{P}_{h^{(1)}}(y=1 \mid x),
\]
\[
\hat{\pi}^{(2)}(x) = \mathbf{P}_{h^{(2)}}(y=1 \mid x).
\]

Given a labelled training set
\(
\{(x_i,y_i)\}_{i=1}^N
\)
and a disjoint validation set
\(
\{(x_i^{\mathrm{val}},y_i^{\mathrm{val}})\}_{i=1}^{N_{\mathrm{val}}}
\),
the training procedure goes as follows: train each base estimator \(h^{(k)}\) on the training set \((X_{\mathrm{train}}, y_{\mathrm{train}})\), optionally using the validation set for early stopping; fit a temperature parameter for each base model on the validation set, obtaining calibrated probabilities \(\hat{\pi}^{(k)}(x)\); see Section~\ref{subsubsec:temp_calib}; using the calibrated probabilities on the validation set, construct targets for the router that indicate where the secondary model outperforms the primary model (details below); train an XGBoost router~\cite{chen2016xgboost} on the validation features and these router targets.

At test time, the router decides for each input whether to use the primary (classical) or secondary (quantum) expert, yielding a mixture-of-experts style architecture~\cite{jordan1994hme} that adaptively exploits the strengths of both models.

\subsubsection{Router training}

Let
\(
\hat{\pi}^{(1)}_i = \hat{p}^{(1)}(x_i^{\mathrm{val}})
\)
and
\(
\hat{\pi}^{(2)}_i = \hat{p}^{(2)}(x_i^{\mathrm{val}})
\)
denote the calibrated probabilities of the primary and secondary models on the validation examples, and let \(y_i^{\mathrm{val}} \in \{0,1\}\) be the corresponding labels.
First, for each expert \(k \in \{1,2\}\) we choose an operating threshold \(\tau^{(k)}\) using Youden's \(J\) statistic \cite{youdenj2005}.  
For a candidate threshold \(\tau\), define the true positive rate and false positive rate of model \(k\) on the validation set as
\[
\mathrm{TPR}^{(k)}(\tau)
= \mathbf{P}\big(\hat{\pi}^{(k)}_i > \tau \,\big|\, y_i^{\mathrm{val}}=1\big),
\]
\[
\mathrm{FPR}^{(k)}(\tau)
= \mathbf{P}\big(\hat{\pi}^{(k)}_i > \tau \,\big|\, y_i^{\mathrm{val}}=0\big),
\]
and Youden's index
\(
J^{(k)}(\tau) = \mathrm{TPR}^{(k)}(\tau) - \mathrm{FPR}^{(k)}(\tau).
\)
We select
\[
\tau^{(k)} = \arg\max_{\tau \in [0,1]} J^{(k)}(\tau),
\]

Define the indicator function of the logical proposition $A$ as
\begin{equation}
    \mathbf{1}\{A\} = \begin{cases}
        1&\text{if $A$ is true},\\
        0&\text{if $A$ is false}.
    \end{cases}
\end{equation}
Then, using these thresholds we convert probabilities into hard predictions
\[
\hat{y}^{(k)}_i = \mathbf{1}\big\{\hat{\pi}^{(k)}_i > \tau^{(k)}\big\},
\qquad k \in \{1,2\},
\]
and construct a binary router target \(z_i \in \{0,1\}\) that indicates where the secondary model is preferred:
\begin{equation}
    z_i =
    \begin{cases}
        1, & \text{if } \hat{y}^{(2)}_i = y_i^{\mathrm{val}} \text{ and } \hat{y}^{(1)}_i \neq y_i^{\mathrm{val}},\\[2pt]
        0, & \text{otherwise.}
    \end{cases}
\end{equation}
In other words, \(z_i=1\) whenever the secondary (quantum) model correctly classifies a validation point that the primary model misclassifies; all other cases, including regions where both models succeed or both fail, are assigned \(z_i=0\). This focuses the router on learning the subset of feature space where the secondary model strictly outperforms the primary model.
The router is implemented as an XGBoost classifier~\cite{chen2016xgboost}.

At test time, the router outputs a probability
\(g_i(x) \in [0,1]\) that a given input lies in a region where the secondary model is preferred.  A user-controlled parameter \(\gamma \in (0,1)\) (the router threshold) then defines a hard routing decision
\[
r(x) = \mathbf{1}\big\{ g_i(x) > \gamma \big\},
\]
and the final combined probability is given by the hard mixture
\begin{equation}
    \hat{p}_{\mathrm{comb}}(x)
    = \big(1 - r(x)\big)\,\hat{\pi}^{(1)}(x)
      + r(x)\,\hat{\pi}^{(2)}(x),
\end{equation}
which reduces to the primary model when \(r(x)=0\) and to the secondary model when \(r(x)=1\). This architecture is closely related to the classical mixture-of-experts framework with a gating network~\cite{jordan1994hme}, and in our experiments yields improved performance and robustness under noisy conditions compared to either expert alone.

\section{Experiments}\label{sec:experiments}
\subsection{European Credit Card Dataset}
We used a data set consisting of European Credit Card transactions \cite{europeancc} from September 2013 by European cardholders. It is representative of real world data because it comes from real world usage and is highly imbalanced, with roughly 0.172\% of the data classified as fraud.


\subsection{Preprocessing}

A majority of the features in the European Credit Card dataset were obfuscated and normalized using PCA to protect the real data that was collected. 
The "Time" in this dataset represents the elapsed time between the transaction and the first transaction in the dataset. Card transactions data are fundamentally time series data and their characteristics can vary with time. Practitioners will sometimes separate the train and test data sets so that no transaction in the test set occurs at a time before any of the transactions in the training set. This is to better mimic the behaviour of the deployed model in production where the system is classifying future transactions only as opposed to past transactions. However, it is quite difficult due to the imbalance of the data to create this scenario and still have sufficient information to properly train the model. This would mean that we could have splits in the data that would not have any fraud for training. Since we opted to use cross-validation in our benchmark, we decided to remove the "Time" column, so we do not have situations where a model trained on future time steps transactions tries to predict past time step.

Due to the intrinsic nature of quantum computing, angle encoding requires features to lie within a specific range, so a MinMax scaler was applied to all features. During training, the majority class was downsampled to achieve a 50/50 balance ratio between the nominal and fraudulent transactions.

No other preprocessing steps were used during model training or validation.

\subsection{Validation and Testing}



\begin{figure}[h!]
 \centering
  \includegraphics[width=1\linewidth]{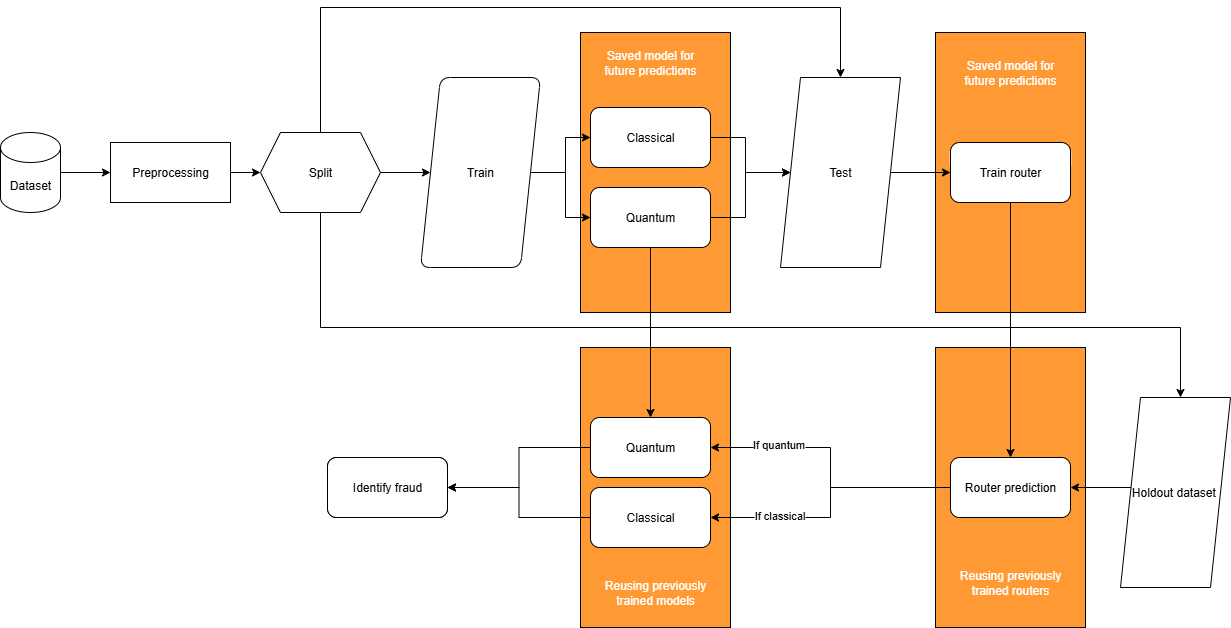}
 \caption{Validation procedure used during experimentation. The dataset is preprocessed using a MinMax scaler to be further split in train, tests, and holdout sets. Train is used to train the model, test is used to train the router, and holdout to evaluate the model's performance.}
 \label{fig:pipeline}
\end{figure}

We utilized stratified cross-validation with 5 folds and 3 repeats and obtained the average and median of Area Under Precision-Recall Curve (AUPRC) and average precision score (AP) metrics to gauge the success of our models. Both metrics captures the same idea which is to summarize the precision-recall curve with its area under the curve, however the former uses the trapezoidal rule while the latter does not. The dataset for each fold was split in  train, analysis, validation, and holdout. The train set remained untouched (other than undersampling it for 50/50 balance ratio), but we further split the test sets in 50/25/25 into validation, analysis, and holdout sets. Validation was used to obtain the optimal threshold of each model, analysis was used to train the router based on the models' predictions, and holdout was treated as unlabelled data for evaluation of the metrics. Figure \ref{fig:pipeline} gives an overall perspective of the framework of the model. 

The hyperparameters selected for the Combined Model were hidden layers of 256, 128, 64 neurons for the autoencoder and 8 neurons for the classification head, 6 qubits for the VQC with 6 layers, $\lambda$ of $0.5$, and batch size of 32. Training and inference were performed using the Pennylane \cite{pennylane} software package.  A The clock-time of the hardware was obtained using OQC's plug-in for the Pennylane software package for 10 different points. The average AUCPR, AP and median AUCPR, AP for our baseline XGBoost and our model with increasing router threshold can be found in table \ref{tab:results-aucpr} and \ref{tab:results-ap}. As can be seen, our model outperforms XGBoost for the credit card dataset on average and on median.


Precision and recall are reported in Tables \ref{tab:results-precion} and \ref{tab:results-recall}. Youden’s J statistic was used to select the operating point for these metrics and suggests that the observed gains in class separability may arise from a trade-off between false positives and fraud detections. For a fixed router threshold, our model consistently achieves higher precision, indicating a reduction in false positives; however, this improvement comes at the cost of lower recall, meaning that fewer fraudulent transactions are detected. In the context of credit card transactions, this trade-off remains valuable for the reasons discussed in the introduction: false positives are often as costly as fraud itself due to their associated fallout.


\begin{table}[ht]
    \centering
    \caption{Table containing the average and median AUCPR of our models obtained from repeated cross validation with 5 folds and three repeats. As indicated, our model achieves comparable or marginally higher AUCPR at the 0.6–0.7 router threshold compared to XGBoost.}
\resizebox{0.7\linewidth}{!}{%
\begin{tabular}{ |p{3cm}||p{3cm}|p{3cm}|  }
 \hline
 \multicolumn{3}{|c|}{AUCPR comparison of models} \\
 \hline
 & Average AUCPR &Median AUCPR\\
 \hline
 XGBoost   & $0.78\pm0.10$&   0.797\\
 This model with 0.5 router threshold&   $0.78\pm0.09$  & 0.799 \\
 This model with 0.6 router threshold &$\mathbf{0.79\pm0.09}$ & \textbf{0.813}\\
 This model with 0.7 router threshold    &$0.79\pm0.08$ &0.813\\
 This model with 0.8 router threshold&   $0.78\pm0.09$  & 0.805\\
 This model with 0.9 router threshold& $0.78\pm0.11$  & 0.803   \\
 \hline
\end{tabular}}

    \label{tab:results-aucpr}
\end{table}

\begin{table}[ht]
    \centering
    \caption{Table containing the average and median Precision of our models obtained from repeated cross validation with 5 folds and three repeats. As indicated, our model marginally higher precision with the threshold defined by Youden's J at any router threshold compared to XGBoost.}
\resizebox{0.7\linewidth}{!}{%
\begin{tabular}{ |p{3cm}||p{3cm}|p{3cm}|  }
 \hline
 \multicolumn{3}{|c|}{Precision comparison of models} \\
 \hline
 & Average Precision & Median Precision\\
 \hline
 XGBoost   & $0.081\pm0.059$&   0.053\\
 This model with 0.5 router threshold&   $\mathbf{0.127\pm0.142}$  &  \textbf{0.058} \\
 This model with 0.6 router threshold &$0.122\pm0.133$ & 0.057\\
 This model with 0.7 router threshold    & $0.117\pm0.127$ & 0.058\\
 This model with 0.8 router threshold&   $0.099\pm0.089$  & 0.058\\
 This model with 0.9 router threshold& $0.092\pm0.076$  & 0.058   \\
 \hline
\end{tabular}}

    \label{tab:results-precion}
\end{table}

\begin{table}[ht]
    \centering
    \caption{Table showing that our model achieves lower recall than XGBoost, indicating a precision/recall trade-off: the hybrid model reduces false positives at the cost of missing a small fraction of additional fraud cases. Which suggests why precision metrics were higher, but class separability ones like AUCPR and AP are only marginally higher.}
\resizebox{0.7\linewidth}{!}{%
\begin{tabular}{ |p{3cm}||p{3cm}|p{3cm}|  }
 \hline
 \multicolumn{3}{|c|}{Recall comparison of models} \\
 \hline
 & Average Recall &Median Recall\\
 \hline
 XGBoost   & $\mathbf{0.934\pm0.051}$&   \textbf{0.933}\\
 This model with 0.5 router threshold&   $0.909\pm0.057$  &  0.931\\
 This model with 0.6 router threshold &$0.913\pm0.053$ & 0.931\\
 This model with 0.7 router threshold    &$0.915\pm0.049$ & 0.931\\
 This model with 0.8 router threshold&   $0.918\pm0.048$  & 0.931\\
 This model with 0.9 router threshold& $0.923\pm0.051$  & 0.931  \\
 \hline
\end{tabular}}

    \label{tab:results-recall}
\end{table}

\begin{table}[ht]
    \centering
    \caption{Table containing the average and median average precision scores (AP) of our models obtained from repeated cross validation with 5 folds and three repeats. The 0.6 router threshold achieves, marginally,
    the highest average AP (0.793) vs. XGBoost (0.770).}
\resizebox{0.7\linewidth}{!}{%
\begin{tabular}{ |p{3cm}||p{3cm}|p{3cm}|  }
 \hline
 \multicolumn{3}{|c|}{Average precision score (AP) comparison of models} \\
 \hline
 & Average AP &Median AP\\
 \hline
 XGBoost   & $0.770\pm0.096$&   0.798\\
 This model with 0.5 router threshold&   $0.785\pm0.085$  & 0.799 \\
 This model with 0.6 router threshold &$\mathbf{0.793\pm0.085}$ & \textbf{0.813}\\
 This model with 0.7 router threshold    &$0.792\pm0.084$ & 0.813\\
 This model with 0.8 router threshold&   $0.785\pm0.086$  & 0.806\\
 This model with 0.9 router threshold& $0.778\pm0.11$  & 0.803   \\
 \hline
\end{tabular}}
    \label{tab:results-ap}
\end{table}

One of the key points raised regarding the limitations of quantum machine learning is the clock-speed of quantum hardware. Due to its infancy and limited accessibility, the use of near-term hardware is often bottlenecked by server, compilation, and execution times. Server time refers to the duration from when a task is received by the cloud to when it is sent to the compiler. Compilation time is the time taken for the compiler to receive the task, optimize the circuit, perform qubit routing, and submit it for execution. Execution time refers to the duration required by the hardware to execute the provided quantum circuit.   

\begin{table}[ht]
    \centering
    \caption{Table containing the average server, compilation, and execution time of OQC's hardware for 10 tasks.}
\resizebox{0.7\linewidth}{!}{%
\begin{tabular}{ |p{3cm}||p{3cm}|p{3cm}|  }
 \hline
 \multicolumn{3}{|c|}{Average Clock-time of the quantum hardware in seconds} \\
 \hline
 Server time & Compilation time & Execution time \\
 \hline
 $0.17\pm0.01$   & $1.92\pm0.11$&   $0.649\pm0.007$\\
 \hline
\end{tabular}}
    \label{tab:clock-time}
\end{table}

Table \ref{tab:clock-time} presents the average server, compilation, and execution times in seconds of Oxford Quantum Circuits (OQC) Toshiko hardware. In the data splits we obtained, the holdout set contained approximately 14,000 points, implying that a fully quantum architecture would require nearly 12 hours for inference alone. Using our proposed method, the worst-case scenario occurred with a router threshold of 0.5 on the first fold of the first split, where approximately $3\%$ of the points were submitted to quantum hardware, resulting in an additional inference overhead of 21 minutes. In the best-case scenario, with a router threshold of 0.9, approximately $1\%$ of the points were submitted to the quantum hardware, adding only 7 minutes to the total inference time.

\section{Discussion and remarks}\label{sec:discussion}


As we have demonstrated, incorporating a gating model within a quantum–classical hybrid architecture can lead to improved overall performance. This approach has meaningful real-world implications for time-sensitive systems, such as those deployed in the financial industry, where latency and reliability are critical, and it is achieved without sacrificing accuracy as measured by AUCPR and average precision (AP) scores. Our model achieved an average AUCPR (AP) of $0.78\pm0.09$ ($0.785\pm0.085$) and a median of 0.799 (0.799) for 0.5 router threshold and an average of $0.79\pm0.09$ ($0.793\pm0.085$) and median of 0.813 (0.813) for 0.6 threshold, compared to the XGBoost baseline, which obtained an average AUCPR (AP) of $0.78\pm0.10$ ($0.770\pm0.096$) and a median of 0.797 (0.798). Our model showed an increase in precision $0.127\pm0.142$ for 0.5 router threshold and $0.122\pm0.133$ for 0.6 router threshold compared to $0.081\pm0.059$ for XGBoost and a reduction in recall from $0.934\pm0.051$ for XGBoost to $0.909\pm0.057$ for our model with router threshold of 0.5 and $0.913\pm0.053$ for router threshold of 0.6. This highlights a possible trade-off between the number of fraud detected and a reduction in false positives.

The inclusion of a dedicated classification head provides an additional degree of flexibility, allowing the system to be tuned for more favourable outcomes and, together with the choice of ansatz, serving as a key differentiator from the original GQC architecture. Furthermore, the use of temperature calibration through Platt scaling facilitates a more effective and stable neural network implementation, improving both the interpretability and robustness of the model’s predictions.

To demonstrate that our method is compatible with current quantum hardware, we show that in the best-performing configuration the model incurs only an additional 7 minutes of inference time for approximately 14,000 data points, compared to nearly 12 hours when a fully quantum model is used alone. This indicates the value of utilising the MoE approach with a fast quantum computing platform such as superconducting circuits: by combining these, we can obtain the fastest possible quantum-enhanced results in low-latency applications such as fraud detection.


\section{Open Question and future outlook}

We believe improvements could be obtained by implementing alternative gating systems by identifying if there is a more suitable MoE strategy available on the literature. Another strategy would be identifying data complexities that can be linked to the success of quantum machine learning models and use it in the gating process.



\section{Acknowledgements}

The authors from OQC would like to thank Aamna Irfan, Daria Van Hende, and Alex Lillistone for their help in building the cloud infrastructure to make this work a reality. We also thank the rest of the DevOps team, especially Luke Batty, Scott Followell, and Tom Winchester, for their endless IT support, as well as Joseph Bilella for his operational and customer-facing management. Finally, we thank Owen Arnold and Peter Leek for their review and suggestions.



\clearpage
\bibliographystyle{IEEEtran}
\bibliography{references}

\end{document}